\documentclass[nanomaterials,article,accept,moreauthors,pdftex]{Definitions/mdpi}

\firstpage{1}
\pubvolume{1}
\issuenum{1}
\articlenumber{0}
\pubyear{2021}
\copyrightyear{2020}
\datereceived{\hl{date}}
\dateaccepted{\hl{date}}
\datepublished{\hl{date}}

\pdfoutput=1

\Title{Simplicity Out of Complexity: Band Structure for W$_{20}$O$_{58}$~Superconductor}
\TitleCitation{Simplicity Out of Complexity: Band Structure for W$_{20}$O$_{58}$ Superconductor}

\Author{A.A. Slobodchikov $^{1}$\orcidA{}, I.A. Nekrasov $^{1}$\orcidC{}, N.S. Pavlov $^1$\orcidD{} and M.M. Korshunov$^{2,}$*\orcidB{}}  

\AuthorCitation{Slobodchikov, A.A.; Nekrasov, I.A.; Pavlov, N.S.; Korshunov, M.M.}
\AuthorNames{A.A. Slobodchikov, I.A. Nekrasov, N.S. Pavlov and M.M. Korshunov}

\address{%
$^{1}$ \quad Institute of Electrophysics, Russian Academy of Sciences, Ural Branch, 620016 Ekaterinburg, Russia; stalfear@gmail.com (A.A.S.); nekrasov@iep.uran.ru (I.A.N.); pavlov@iep.uran.ru (N.S.P.)\\ 
$^{2}$ \quad Kirensky Institute of Physics, Federal Research Center KSC SB RAS, Akademgorodok, \mbox{660036 Krasnoyarsk, Russia}} 

\corres{Correspondence: mkor@iph.krasn.ru}

\abstract{The band structure, density of states, and the Fermi surface of a recently discovered superconductor, oxygen-deficient tungsten oxide WO$_{2.9}$ that is equivalent to W$_{20}$O$_{58}$, is studied within the density functional theory (DFT) in the generalized gradient approximation (GGA). Here we show that despite the extremely complicated structure containing 78 atoms in the unit cell, the low-energy band structure is quite feasible. Fermi level is crossed by no more than 10 bands per one spin projection (and even 9 bands per pseudospin projection when the spin-orbit coupling is considered) originating from the t$_{2g}$ 5$d$-orbitals of tungsten atoms forming zigzag chains. These bands become occupied because of the specific zigzag octahedra distortions. To demonstrate the role of distortions, we compare band structures of W$_{20}$O$_{58}$ with the real crystal structure and with the idealized one. We also propose a basis for a minimal low-energy tight-binding model for W$_{20}$O$_{58}$.}

\keyword{band structure; DFT; superconductivity}

\begin{document}

\section{Introduction}
The discovery of a new type of superconductor is always exciting since it promotes a novel insight into the understanding of such a basic phenomenon. It is especially evident in the unconventional and high-$T_c$ superconductors such as cuprates~\cite{bednorz-muller} and Fe-based materials~\cite{y_kamihara_08}. Recent discovery of superconductivity in a tungsten oxide WO$_{2.9}$ is not an exception~\cite{Shengelaya2020}. Structure and electrical properties of tungsten trioxide WO$_{3}$ and oxygen-deficient tungsten oxides WO$_{3-x}$ were thoroughly studied quite long ago~\cite{Bursill1972,Sahle1983}. Tungsten oxides are known for their thermoelectric applications~\cite{Kieslich2016} that makes them the distant relatives to the other superconductor, water-intercalated sodium cobaltate Na$_x$CoO$_2\cdot$H$_2$O~\cite{Terasaki1997,Kawata1999,Takada2003,IvanovaKorshunov2009eng}. Another similarity with sodium cobaltates, cuprates, and iron-based materials arise from the partially filled $d$-orbitals of W involved in the conductivity, i.e., the stoichiometric material WO$_3$ has an empty $d$-shell with tungsten W$^{6+}$ in $5d^0$ configuration.~Oxygen deficiency induces W$^{5+}$ ions with the $5d^1$ configuration.
Observation of superconductivity in twin-walls of WO$_{3-x}$~\cite{Aird1998}, thin films~\cite{Kopelevich2015}, and in WO$_3$ with the surface composition Na$_{0.05}$WO$_3$~\cite{Reich1999} led to the proposal of the possible superconductivity in WO$_{3-x}$~\cite{Shengelaya2019} and a consequent discovery of it in WO$_{2.9}$ with $T_c=80$K and with $T_c=94$K after the lithium intercalation~\cite{Shengelaya2020} that provides even stronger electron doping.

There are quite a few first-principles studies of WO$_3$ regarding its electronic structure~\cite{Hamdi2016,Wijs1999}, role of oxygen vacancies~\cite{Wang2011_wo3_prb,Mehmood2016,Migas2010_1}, and cation doping~\citep{Walkingshaw2004,Tosoni2014,Huda2008} (for more references see review~\cite{Hamdi2016}) but not as much of Magn\'{e}li phases of the tungsten oxide. As far as we know only the work of Migas et al.~\cite{Migas2010_2} contains results for the band structure for all Magn\'{e}li phases (see Figure~\ref{fig:migas}), though it does not include effects of the spin-orbit coupling. In~that paper, the authors have shown that Magn\'{e}li phases of tungsten oxides (W$_{32}$O$_{84}$, W$_3$O$_8$, W$_{18}$O$_{49}$, W$_{17}$O$_{47}$, W$_5$O$_{14}$, W$_{20}$O$_{58}$, and W$_{25}$O$_{73}$) demonstrate the metal-like properties. One of the main features of the band structures is an energy gap of about 1~eV right below the Fermi level. A trend could be traced: the closer the stoichiometry of WO$_x$ to WO$_3$, the smaller the charge carrier concentration. Also, the spin-polarized calculations for the W$_{18}$O$_{49}$ phase does not reveal any magnetic moment and it is claimed that this is valid for the other phases~\cite{Migas2010_2}.

Here we take further steps toward the understanding of the WO$_{2.9}$ properties. Specifically, by means of density functional theory (DFT), we explore the orbital composition of bands crossing the Fermi level and give detailed explanation of why exactly these orbitals are occupied. We also discuss the topology of the Fermi surface and the role of the spin-orbit coupling. To see the influence of crystal structure distortions on the band structure, comparison with the idealized crystal structure W$_{20}$O$_{58}$ bands is performed. The unit cell of W$_{20}$O$_{58}$ contains 78 ions and produces an extremely complicated band structure. We show that one can, however, formulate a rather simple low-energy band model that includes just a few parabolic bands of the $d_{xz}$ and $d_{yz}$ orbital character.
\begin{figure}[H]

(a)\includegraphics[scale=0.31]{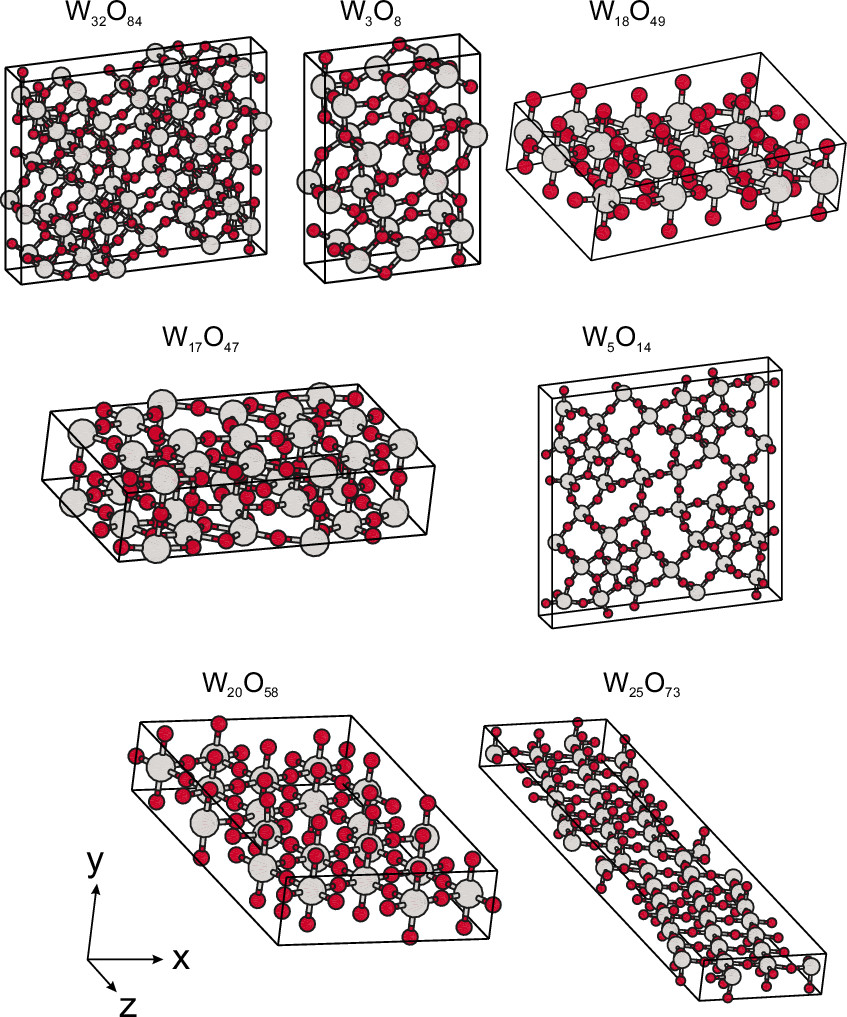}
\caption{\textit{Cont.}}
\end{figure}

\begin{figure}[H]\ContinuedFloat
(b)\includegraphics[scale=0.22]{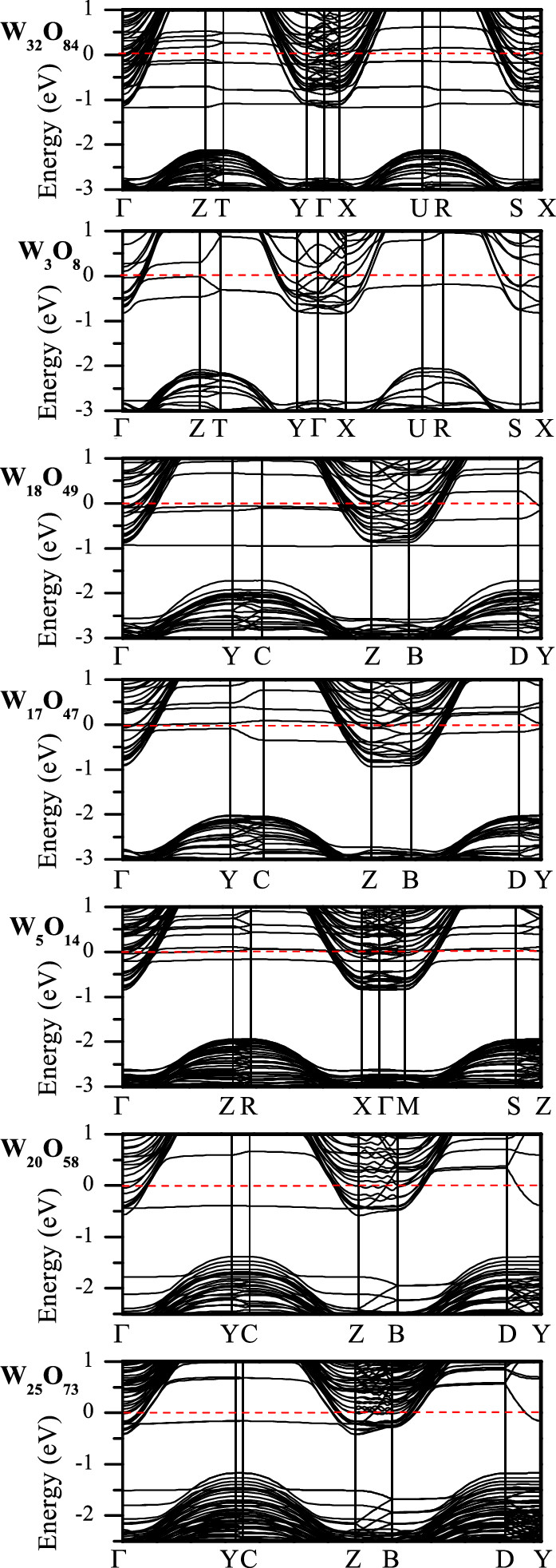}
\caption{Crystal structure of Magn\'{e}li phases (\textbf{a}) and corresponding band structures (\textbf{b}). Reprinted from~\cite{Migas2010_2}, with the permission of AIP Publishing, 2010.\label{fig:migas}}
\end{figure} 

\vspace{-18pt}
\section{Crystal Structure and Calculation Details}

\begin{figure}[H]

\includegraphics[width=1\linewidth]{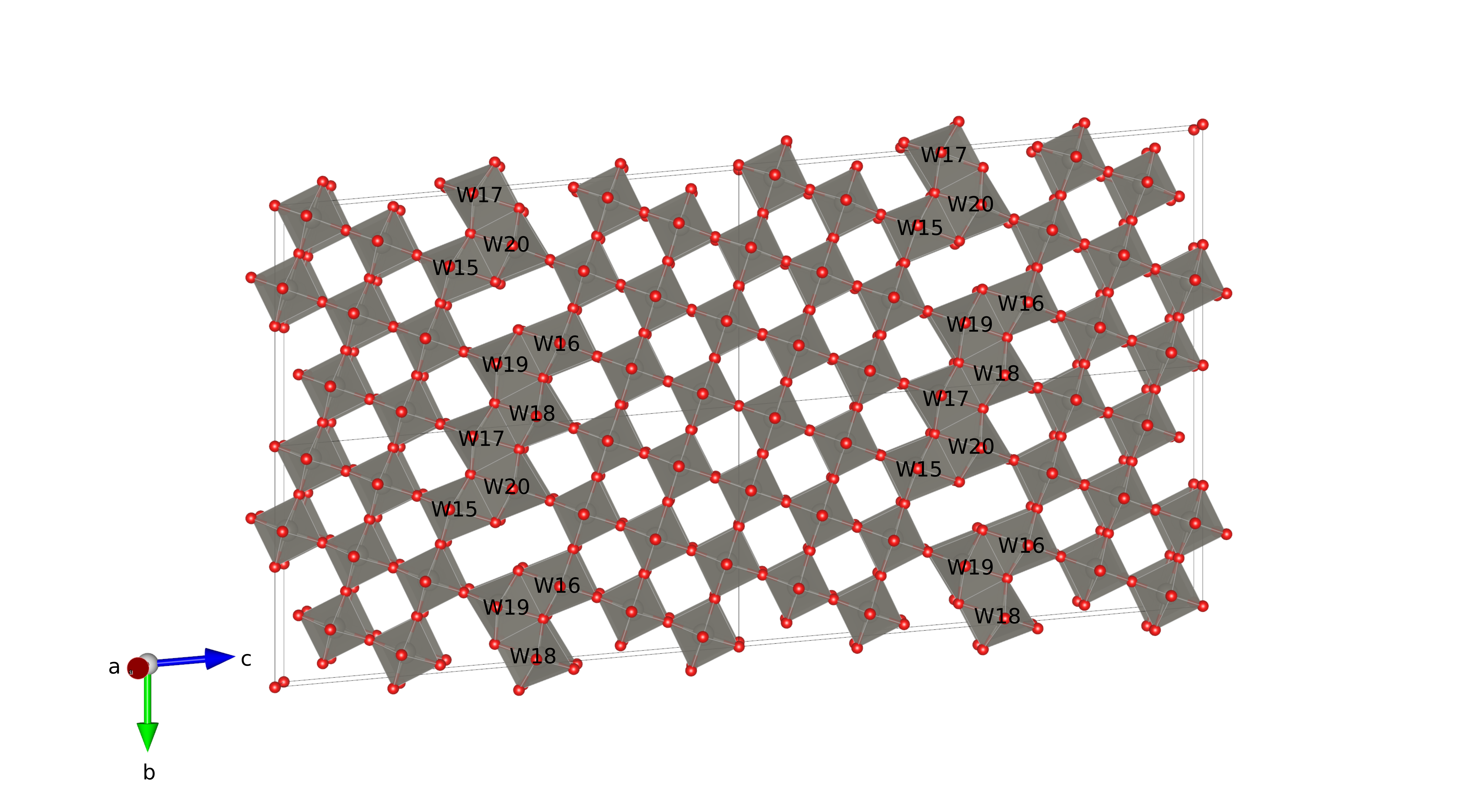}
\caption{Crystal structure of the W$_{20}$O$_{58}$ supercell. Tungsten atoms located along the zigzag stripes are labeled. \label{fig:structure}}
\end{figure}

\textls[-15]W$_{20}$O$_{58}$ belongs to the family of the Magn\'{e}li-type oxides with the general formula W$_n $O$_{3n-2}$~\cite{Bursill1972}. The space group is a $P2$/$m$:$b$ with the unique axis $b$. The lattice parameters are the following: $a = 3.78$ \AA, $b = 12.1$ \AA, $c = 23.39$ \AA, $\beta = 95^{\circ}$~\cite{Magneli1949}. In Figure~\ref{fig:structure} we show the W$_{20}$O$_{58}$ supercell ($2 \times 2 \times 2$).

The elementary unit cell is quite large and contains 20 tungsten atoms and 58 oxygens making it a kind of a nano-object.~Complexity of WO$_{2.9}$ crystal structure arises from the oxygen vacancies ordering.~The main motif of the crystal structure is distorted (with respect to the ideal ones) WO$_6$ octahedra, where the tungsten atom in the center is coordinated by the six oxygen atoms located in the vertices. The entire structure may be described as consisting of two parts: blocks of the corner-sharing octahedra located between the zigzag stripes and blocks of the edge-sharing octahedra located along the stripes. Later builds the stripe-like structures. Since the superconducting volume fraction is about 20\%, the key role here may be played by these structures~\cite{Shengelaya2020}.

Interestingly, corner-shared octahedra in a sense of atomic positions are not far away from the ideal ones.~On the other side, edge-shared octahedra are rather specifically distorted in the (100) plane. Specifically, one of W--O 
bonds in the basal plane of the zigzag octahedra is quite elongated up to 2.2--2.3 \AA~while other W--O bonds are about 1.9 \AA~that is similar to those in the corner-shared octahedra. Because of that, as we shall see below, tungsten atoms located along the zigzag stripe (labeled as W15--W20 in Figure~\ref{fig:structure}) form the band structure near the Fermi level. 

{To calculate the band structure, the density of states (DOS), and the Fermi surface we use DFT with all-electron full-potential linearized augmented-plane wave (LAPW) implemented via the Elk code~\cite{elk} together with the generalized gradient approximation (GGA)~\cite{jperdew96}.~Spin-orbit coupling (SOC) was included within the fully relativistic calculation~scheme.}

All calculations were converged self-consistently on a grid of $8 \times 8 \times 8$ \textbf{k}-points in the irreducible Brillouin zone. Due to the size of the system, the calculations are rather costly. Therefore, we first calculated only the band structure for several grid sizes and found that results for $6 \times 6 \times 6$ and $8 \times 8 \times 8$ grids are almost the same. Thus, we can confidently use the $8 \times 8 \times 8$ \textbf{k}-points grid for all further calculations. In the right side of Figure~\ref{fig:dos_wide_bands}, 
we show the Brillouin zone with the \textbf{k}-path used in the band structure analysis. High-symmetry \textbf{k}-points were selected according to the SeeK-path tool~\cite{Hinuma2017}. 

\section{Results and Discussion}

\begin{figure}[H]

 \includegraphics[width=0.77\linewidth]{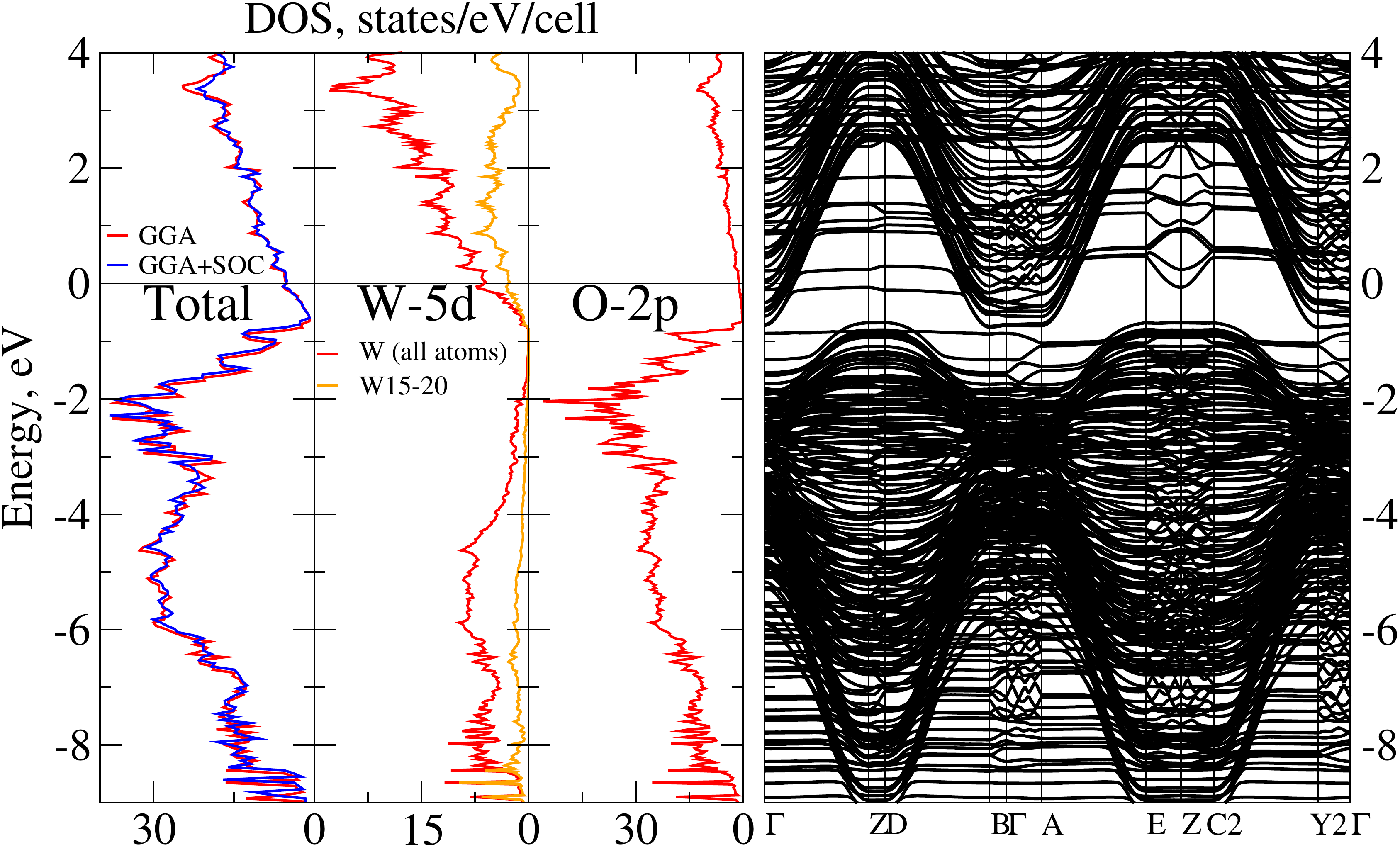}
 \includegraphics[width=0.18\linewidth]{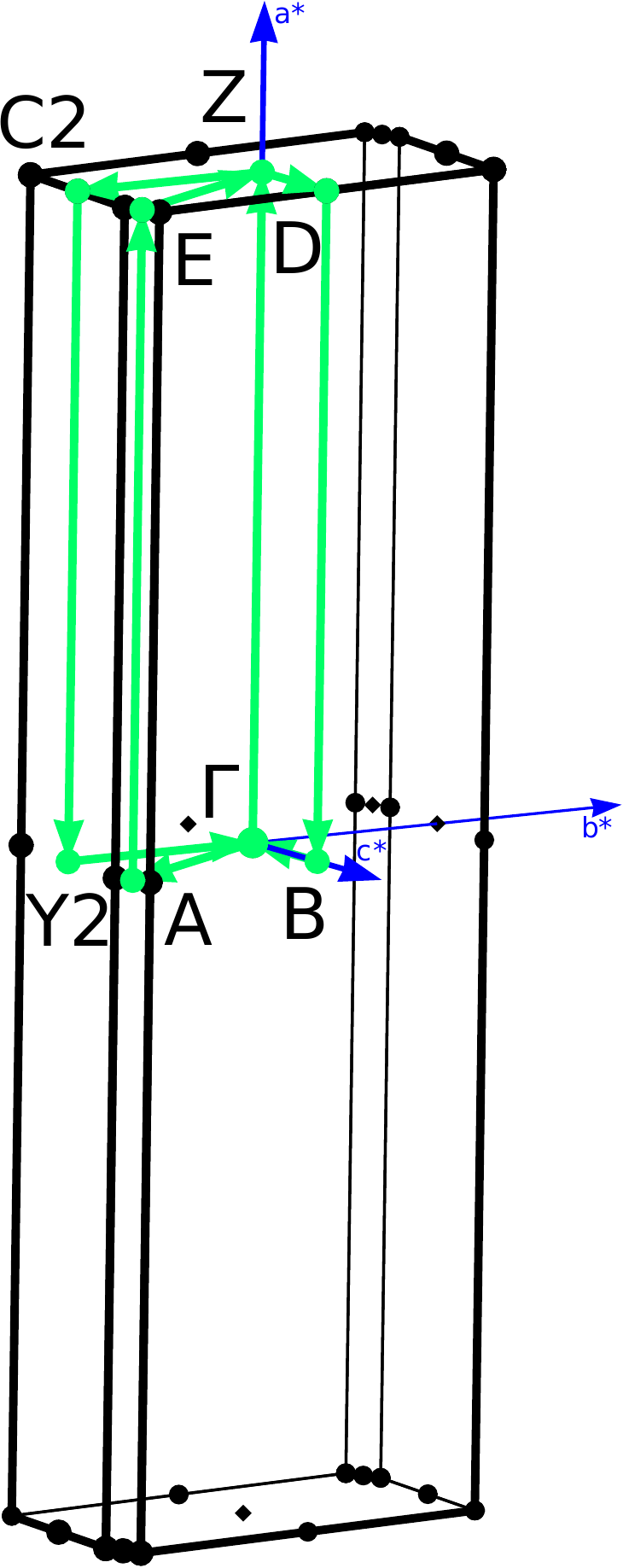}
 \caption{Left: DFT-calculated DOS for W$_{20}$O$_{58}$ --- total DOS with and without SOC (left section), DOS for tungsten atoms contained in the layers of the edge-shared octahedra (middle section), and DOS for oxygen atoms (right section). Center: DFT-calculated band structure in a wide energy range without SOC. Right: Brillouin zone for W$_{20}$O$_{58}$ with the $P2$/$m$:$b$ space group. Zero corresponds to the Fermi level. \label{fig:dos_wide_bands}}
\end{figure}

DOS and the band structure in a wide energy range for W$_{20}$O$_{58}$ are shown in \mbox{Figure~\ref{fig:dos_wide_bands}}. Band structure clearly shows a small gap just above $-1$~eV. As is evident from DOS, the states under the gap originates mostly from oxygen while the states above the gap are mostly from tungsten. This comes from the absence of two oxygen atoms in  W$_{20}$O$_{58}$ as compared to the WO$_{3}$ (equivalent to W$_{20}$O$_{60}$) band insulator system. Oxygen deficiency provides four unbound electrons which occupy the W-5$d$ states. It can be checked by the integration of the total DOS in the energy interval from $-0.8$~eV to $0$ that gives nearly four electrons. Please note that almost three of these electrons belong to the tungsten atoms W15--W20 located along the zigzag stripe in the crystal structure (see Figure~\ref{fig:structure}) while the rest of tungsten atoms have practically empty 5$d$ shells. Those four extra electrons should occupy the $t_{2g}$ orbitals, which are lowest in energy of the W-5$d$ manifold. Introduction of SOC does not lead to any notable changes in the total DOS.

\begin{figure}[H]

 \includegraphics[width=0.9\linewidth]{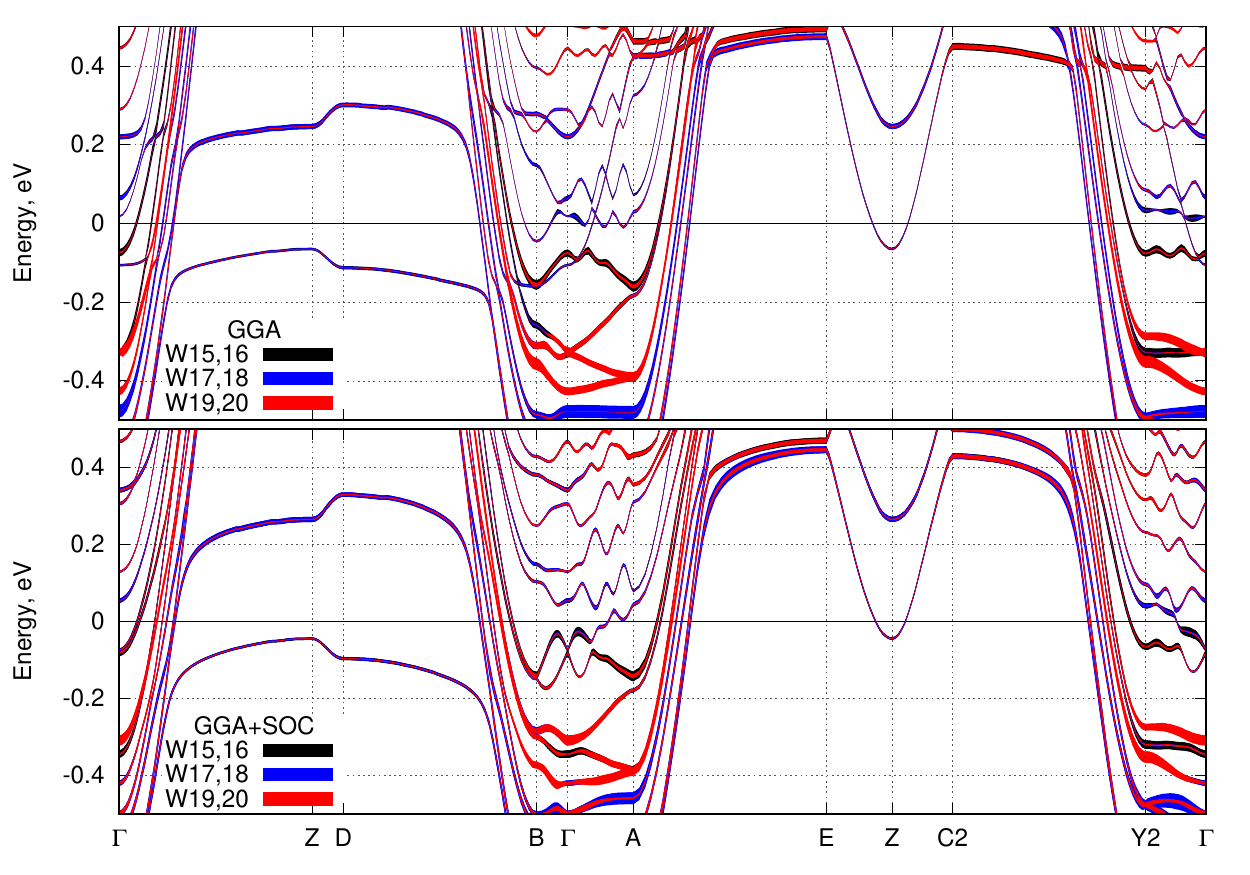}
 \caption{DFT-calculated band structure without SOC (top) and with SOC (bottom). The orbital character of the tungsten atoms W15--W20 are shown by different colors and the width of each curve is proportional to the contribution of the corresponding orbital. Zero corresponds to the Fermi level. \label{fig:fatbands}}
\end{figure}

To demonstrate the importance of the tungsten atoms W15--W20, we plot their contributions to the bands near the Fermi level in Figure~\ref{fig:fatbands}. Contributions from other tungsten atoms near the Fermi level are almost negligible. SOC does not lead to any notable changes of the bands near the Fermi level. However, the spin-orbital coupling leads to the disappearance of the electron pocket around the $B$ point and the disappearance of two smaller electron {pockets in the} $\Gamma-A$ direction.

On the grounds of the calculated band structure and the analysis of the orbital contribution, we propose the basis for the minimal low-energy  tight-binding model. \mbox{In Figures~\ref{fig:dos_wide_bands} and ~\ref{fig:fatbands}}, one can see the bands crossing the Fermi level are pretty much parabolic ones. Analysis of these parabolic bands shows that the $d_{xz}$ and $d_{yz}$ orbitals of the tungsten W-5$d$ $t_{2g}$ manifold of W15--W20 atoms give the dominant contribution near the Fermi level. The $d_{xz}$ and $d_{yz}$ orbitals become occupied because of the specific distortions of the WO$_6$ octahedra within the zigzag structure described above. Since one of the basal oxygens is quite far from the central tungsten atom, those $d$-orbitals are less hybridized with the O-2$p$ states and thus are shifted to lower energies, and consequently get electrons on them. Despite that the other corner-shared octahedra are also distorted (although not that much), splitting of their tungsten's $t_{2g}$ orbitals is not strong enough to allow for the occupation. Therefore, the minimal low-energy model should include just a few parabolic bands of the $d_{xz}$ and $d_{yz}$ orbital character.
	
Let us also note that these parabolic bands have a quite strong dispersion along the $z$-axis (Brillouin zone directions $\Gamma-Z$, $A-E$, etc.). In particular, they form `bell'-shaped bands in the $\Gamma-Z-B$ direction from $-1$~eV to $3$~eV. The top of the `bell' is quite flat leading to the bunch of flat, or `quasi-one-dimensional', bands seen in $\Gamma-Z$, $D-B$, $A-E$, and $C2-Y2$ directions, although the W$_{20}$O$_{58}$ is essentially a three-dimensional system. 
There are also a few bands with the dispersion in the $k_x-k_y$ plane, see, e.g., direction $E-Z-C2$. As for the `wavy' bands in the planar $B-\Gamma-A$ and $Y2-\Gamma$ directions, there is a simple explanation of their \textbf{k}-dependence. Due to the large number of atoms in the unit cell, there is a strong folding of the Brillouin zone in the $k_x-k_y$ plane. Thus, the bands in the $B-\Gamma-A$ and $Y2-\Gamma$ directions heavily intersect each other and a lot of gaps opens at the intersection points.
Parabolic nature of the bands can be traced by the bare eye.
By coincidence, because of the folding, there are just a few Fermi level crossings in those~directions.

\begin{figure}[H]

 (a)\includegraphics[width=0.2\linewidth]{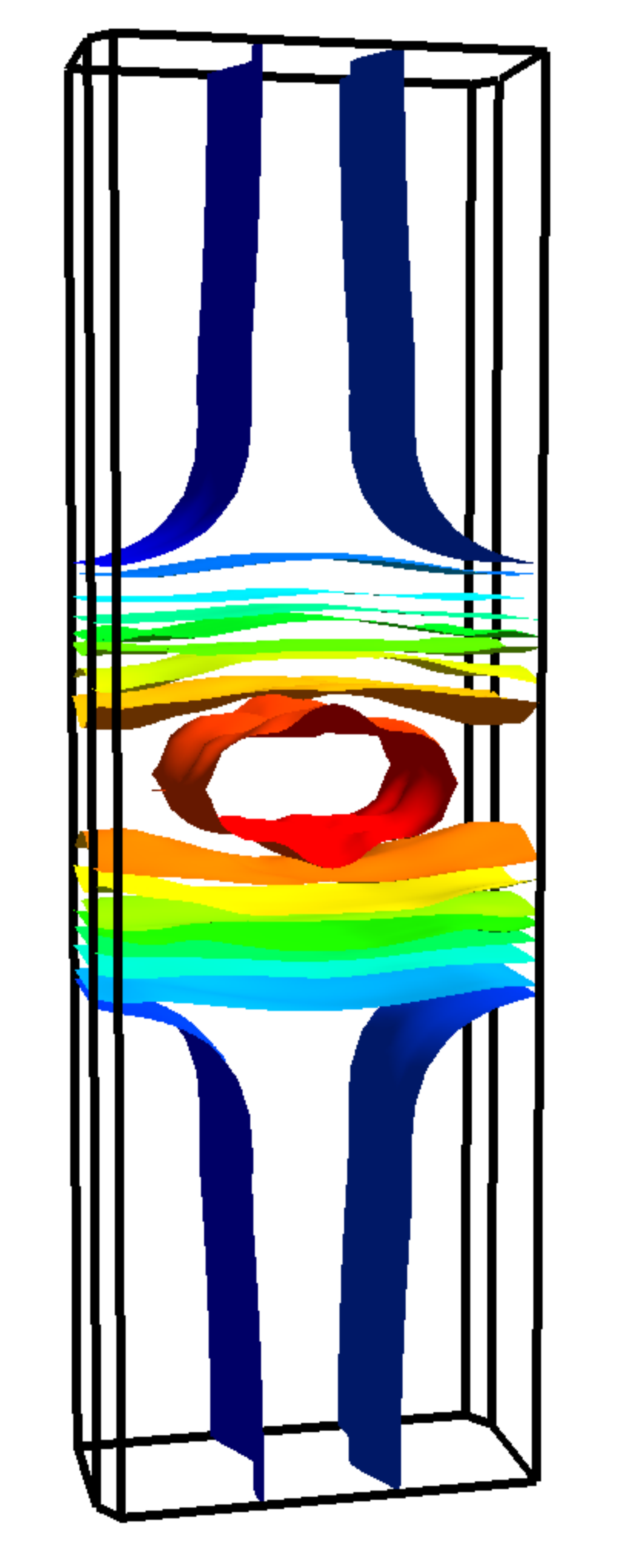}
 (b)\includegraphics[width=0.16\linewidth]{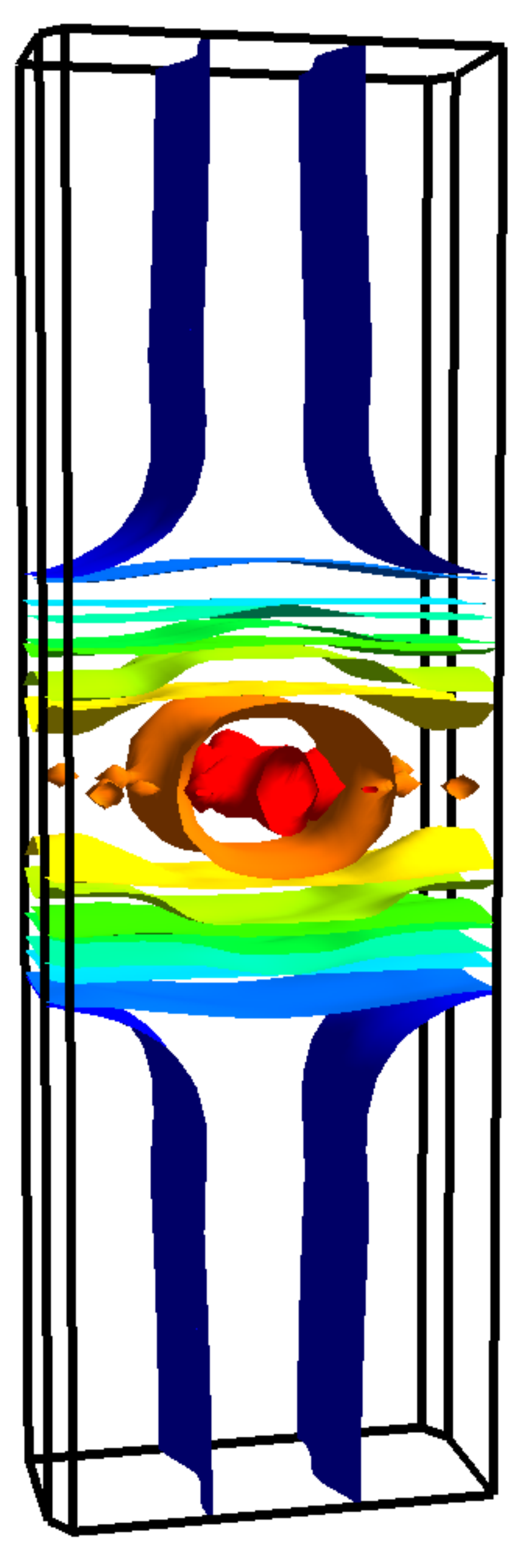}
 \caption{Comparison of the DFT-calculated Fermi surface for W$_{20}$O$_{58}$ with SOC (\textbf{a}) and without SOC (\textbf{b}). \label{fig:fermi_surfaces}}
\end{figure}

GGA calculated Fermi surfaces with and without SOC are shown in Figure~\ref{fig:fermi_surfaces}a and Figure~\ref{fig:fermi_surfaces}b, respectively. There is a bunch of one-dimensional Fermi surface sheets parallel to the $\Gamma-A-B$ plane and a couple of  two-dimensional Fermi surface sheets around $\Gamma$ and $C2$ points. Again, once the SOC is switched on, the electron pocket around the $B$ point and two small electron pockets in the $\Gamma-A$ direction disappear. Therefore, the SOC makes the band structure just a bit simpler. Yet there are not so many bands crossing the Fermi level and, when the spin-orbit coupling is considered, the Fermi surface consists of 18~sheets with 14 of them with a quasi-one-dimensional character and the rest with a two-dimensional character.

\begin{figure}[H]

\includegraphics[width=0.89\linewidth]{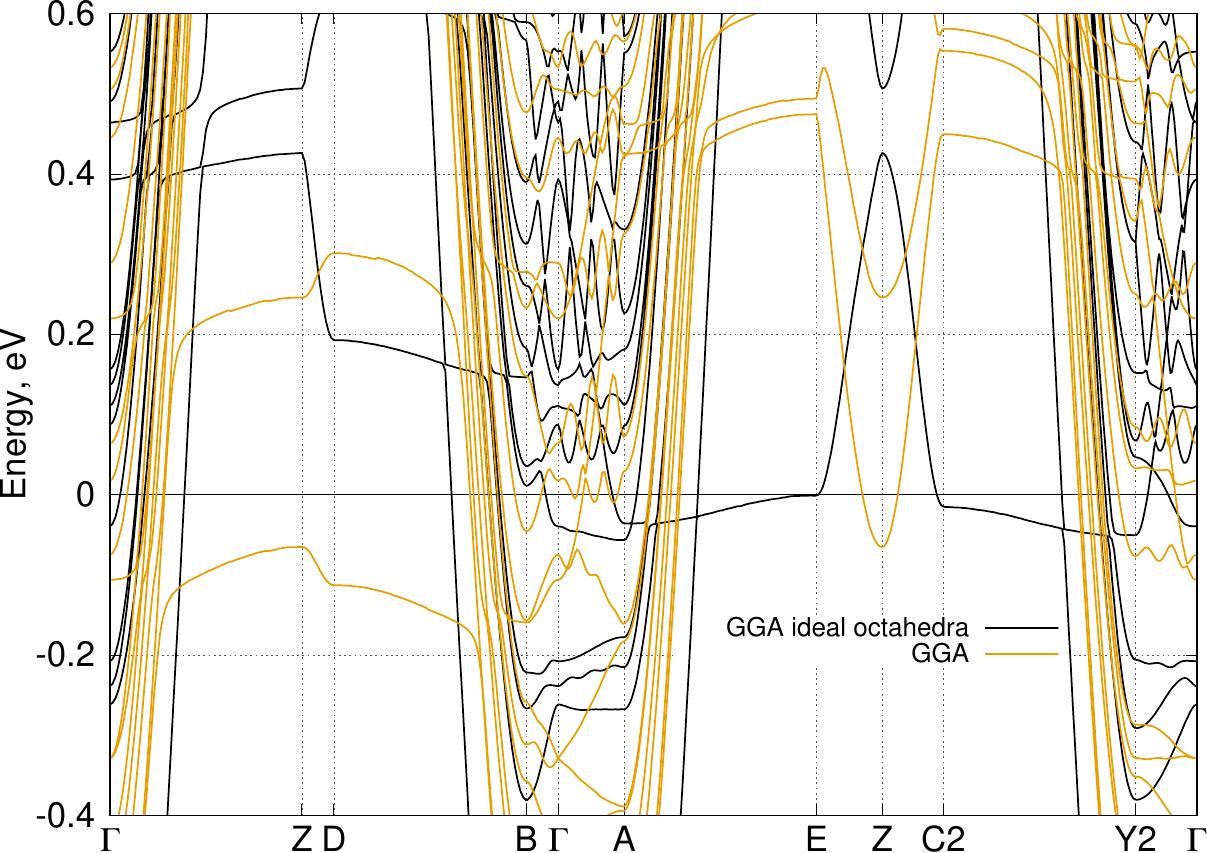}
\caption{Comparison of the DFT-calculated band structure for the real W$_{20}$O$_{58}$ without SOC (orange) and W$_{20}$O$_{58}$ with the ideal octahedra (black)~\cite{Korshunov2021}. Zero corresponds to the Fermi level. \label{fig:bands_real+ideal}} 
\end{figure}

Additional complexity of the W$_{20}$O$_{58}$ system comes from the structure of the WO$_6$ octahedra, i.e., they are all distorted and the O-O bond length ranges from 2.63~\AA~to 2.72~\AA. To see what would happen for an idealized structure, we have turned all WO$_6$ octahedra into ideal ones with the average O-O bond length equal to 2.68~\AA~\cite{Korshunov2021}. Thus, the bases of all ideal octahedra create the ``square lattice''. In Figure~\ref{fig:bands_real+ideal}, we present the comparison of band structures for the real W$_{20}$O$_{58}$ considered here and for the model system with the ideal octahedra from Ref.~\cite{Korshunov2021}. In the latter case, the main noticeable feature is the appearance of the flat bands near the Fermi level in the $A-E$ direction. Those bands originate from the $d$-states of tungsten atoms located along the zigzag stripe, see Figure~\ref{fig:structure}. On the one hand, there is indeed some simplification of the electronic structure: only 6 bands per one spin projection is crossing the Fermi level in contrast to 10 for the real system without SOC. On the other hand, the Fermi surface becomes more ``three-dimensional'' for the idealized structure, i.e., the central part is a closed surface contrary to the distorted cylinder in the real system, see Figure~\ref{fig:fermi_surfaces}.

\section{Conclusions}

We have studied the electronic structure of the non-stoichiometric material WO$_{2.9}$ that is equivalent to W$_{20}$O$_{58}$ via the state-of-art density functional theory in the generalized gradient approximation (GGA). The material has a sophisticated unit cell containing 78~ions thus producing an extremely complicated band structure. Bands crossing the Fermi level, however, originate mostly from the W-5$d$-orbitals of the small number of tungsten atoms forming the zigzag stripe pattern in the crystal structure and have predominantly  W-5$d_{xz}$ and W-5$d_{yz}$ character. It is the reason the Fermi surface is quite simple. Indeed, it consists of five two-dimensional and several quasi-one-dimensional sheets. Thus, we conclude that the electronic properties of WO$_{2.9}$ may be governed by the small number of W-5$d$-bands. This leads to a rather simple minimal model consisting of parabolic bands with $d_{xz}$ and $d_{yz}$ character. If the observed superconductivity involves electrons from these bands, it may explain the smallness of the experimentally estimated superconducting volume fraction.

Comparison of band structures for the real W$_{20}$O$_{58}$ with and without SOC and for the model system with the ideal WO$_6$ octahedra gives an unexpected result---the more complex the crystal structure, the simpler the Fermi surface. In particular, in the system with the ideal WO$_6$ octahedra, the Fermi surface contains a few three-dimensional closed surfaces and a several two-dimensional ones. In the real system without the SOC, we see several two-dimensional, few quasi-one-dimensional, and only one closed three-dimensional surface. In addition, in the system with SOC, there are only a few two-dimensional and several quasi-one-dimensional Fermi surface sheets. This makes the real WO$_{2.9}$ system a close relative to other quasi-two-dimensional systems demonstrating an unconventional superconductivity. 

As for the possible mechanism of high-$T_c$ superconductivity, the conventional, electron-phonon interaction-driven, BCS mechanism is expected to give much lower $T_c$'s than those observed experimentally~\cite{Shengelaya2020} because of the relatively low density of states at the Fermi level. On the other hand, similarity of such gross features, such as the quasi-two-dimensional Fermi surface consisting of several sheets formed by the W-5$d_{xz,yz}$-orbitals, between W$_{20}$O$_{58}$ and iron-based materials, as well as high-$T_c$ cuprates, may point towards the unconventional mechanism of Cooper pairing. In particular, multiband nature of the Fermi surface may result in the enhanced spin/orbital interband fluctuations, which are the main candidates for the `pairing glue' in Fe-based superconductors~\cite{HirschfeldKorshunov2011,Kontani,Korshunov2014eng}.

Before proceeding to the superconductivity calculations, one must understand the effect of electronic correlations always present to some extent in transition metal compounds. First-principle scheme LDA+U can be applied to magnetic systems that is not the case of W$_{20}$O$_{58}$. Application of the LDA+DMFT approach here runs into difficulties because the structure of the unit cell is extremely complicated. Model approach seems to be the most promising and represents the next step to be done in the further studies.
\vspace{6pt}



\authorcontributions{Conceptualization, M.M.K. and I.A.N.; calculations, A.A.S. and N.S.P.; writing, M.M.K., I.A.N., and A.A.S.; funding acquisition, I.A.N. and M.M.K. All authors have read and agreed to the published version of the manuscript.}

\funding{This work was supported in part by RFBR grants No. 18-02-00281, 20-02-00011 (IAN, NSP, AAS), by RFBR and Government of Krasnoyarsk Territory and Krasnoyarsk Regional Fund of Science to the Research Projects ``Electronic correlation effects and multiorbital physics in iron-based materials and cuprates'' grant No. 19-42-240007 (MMK), and by the Program of Ministry of Education and Science of the Russian Federation No. 2020-1902-01-239.
NSP work was also supported in part by the President of Russia grant for young scientists No. MK-1683.2019.2.
}


\acknowledgments{We would like to thank S.G. Ovchinnikov and M.V. Sadovskii for useful discussions. The DFT/GGA computations were performed at ``URAN'' supercomputer of the Institute of Mathematics and Mechanics of the RAS Ural Branch.}

\conflictsofinterest{The authors declare no conflict of interest.}

\end{paracol}
\reftitle{References}


\begin{thebibliography}{999}

\end{thebibliography}


\begin{thebibliography}{999}

\bibitem[Bednorz and M{\"u}ller(1986)]{bednorz-muller}
Bednorz, J.G.; M{\"u}ller, K.A.
\newblock {Possible high T$_c$ superconductivity in the Ba-La-Cu-O system}.
\newblock {\em Z. Phys. B Condens. Matter} {\bf 1986}, {\em
  64},~189--193.
\newblock
  doi:{\changeurlcolor{black}\href{https://doi.org/10.1007/BF01303701}{\detokenize{10.1007/BF01303701}}}.

\bibitem[Kamihara \em{et~al.}(2008)Kamihara, Watanabe, Hirano, and
  Hosono]{y_kamihara_08}
Kamihara, Y.; Watanabe, T.; Hirano, M.; Hosono, H.
\newblock Iron-Based Layered Superconductor La[O$_{1-x}$F$_x$]FeAs
  ($x$ = 0.05 -- 0.12) with $T_c = 26$ K.
\newblock {\em J. Am. Chem. Soc.} {\bf 2008}, {\em
  130},~3296--3297.
\newblock
  doi:{\changeurlcolor{black}\href{https://doi.org/10.1021/ja800073m}{\detokenize{10.1021/ja800073m}}}.

\bibitem[Shengelaya \em{et~al.}(2020)Shengelaya, Conder, and
  M{\"u}ller]{Shengelaya2020}
Shengelaya, A.; Conder, K.; M{\"u}ller, K.A.
\newblock Signatures of Filamentary Superconductivity up to 94 K in Tungsten
  Oxide WO$_{2.90}$.
\newblock {\em J.~Supercond. Nov. Magn.} {\bf 2020},
  {\em 33},~301--306.
\newblock
  doi:{\changeurlcolor{black}\href{https://doi.org/10.1007/s10948-019-05329-9}{\detokenize{10.1007/s10948-019-05329-9}}}.

\bibitem[Bursill and Hyde(1972)]{Bursill1972}
Bursill, L.; Hyde, B.
\newblock CS families derived from the ReO$_3$ structure type: An electron
  microscope study of reduced WO$_3$ and related pseudobinary systems.
\newblock {\em J. Solid State Chem.} {\bf 1972}, {\em 4},~430--446.
\newblock
  doi:{\changeurlcolor{black}\href{https://doi.org/https://doi.org/10.1016/0022-4596(72)90159-4}{\detokenize{10.1016/0022-4596(72)90159-4}}}.

\bibitem[Sahle and Nygren(1983)]{Sahle1983}
Sahle, W.; Nygren, M.
\newblock Electrical conductivity and high resolution electron microscopy
  studies of WO$_{3-x}$ crystals with $0 \leq x \leq 0.28$.
\newblock {\em J. Solid State Chem.} {\bf 1983}, {\em 48},~154--160.
\newblock
  doi:{\changeurlcolor{black}\href{https://doi.org/https://doi.org/10.1016/0022-4596(83)90070-1}{\detokenize{10.1016/0022-4596(83)90070-1}}}.

\bibitem[Kieslich \em{et~al.}(2016)Kieslich, Cerretti, Veremchuk, Hermann,
  Panth{\"o}fer, Grin, and Tremel]{Kieslich2016}
Kieslich, G.; Cerretti, G.; Veremchuk, I.; Hermann, R.P.; Panth{\"o}fer, M.;
  Grin, J.; Tremel, W.
\newblock A chemists view: Metal oxides with adaptive structures for
  thermoelectric applications.
\newblock {\em Phys. Status Solidi} {\bf 2016}, {\em 213},~808--823.
  doi:{\changeurlcolor{black}\href{https://doi.org/10.1002/pssa.201532702}{\detokenize{10.1002/pssa.201532702}}}.

\bibitem[Terasaki \em{et~al.}(1997)Terasaki, Sasago, and
  Uchinokura]{Terasaki1997}
Terasaki, I.; Sasago, Y.; Uchinokura, K.
\newblock Large thermoelectric power in ${\mathrm{NaCo}}_{2}{\mathrm{O}}_{4}$
  single crystals.
\newblock {\em Phys. Rev. B} {\bf 1997}, {\em 56},~R12685--R12687.
\newblock
  doi:{\changeurlcolor{black}\href{https://doi.org/10.1103/PhysRevB.56.R12685}{\detokenize{10.1103/PhysRevB.56.R12685}}}.

\bibitem[Kawata \em{et~al.}(1999)Kawata, Iguchi, Itoh, Takahata, and
  Terasaki]{Kawata1999}
Kawata, T.; Iguchi, Y.; Itoh, T.; Takahata, K.; Terasaki, I.
\newblock Na-site substitution effects on the thermoelectric properties of
  ${\mathrm{NaCo}}_{2}{\mathrm{O}}_{4}$.
\newblock {\em Phys. Rev. B} {\bf 1999}, {\em 60},~10584--10587.
\newblock
  doi:{\changeurlcolor{black}\href{https://doi.org/10.1103/PhysRevB.60.10584}{\detokenize{10.1103/PhysRevB.60.10584}}}.

\bibitem[Takada \em{et~al.}(2003)Takada, Sakurai, Takayama-Muromachi, Izumi,
  Dilanian, and Sasaki]{Takada2003}
Takada, K.; Sakurai, H.; Takayama-Muromachi, E.; Izumi, F.; Dilanian, R.A.;
  Sasaki, T.
\newblock Superconductivity in two-dimensional CoO$_2$ layers.
\newblock {\em Nature} {\bf 2003}, {\em 422},~53--55.

\bibitem[Ivanova \em{et~al.}(2009)Ivanova, Ovchinnikov, Korshunov, Eremin, and
  Kazak]{IvanovaKorshunov2009eng}
Ivanova, N.B.; Ovchinnikov, S.G.; Korshunov, M.M.; Eremin, I.M.; Kazak, N.V.
\newblock Specific features of spin, charge, and orbital ordering in
  cobaltites.
\newblock {\em Physics-Uspekhi} {\bf 2009}, {\em 52},~789--810.
\newblock
  doi:{\changeurlcolor{black}\href{https://doi.org/10.3367/UFNe.0179.200908b.0837}{\detokenize{10.3367/UFNe.0179.200908b.0837}}}.

\bibitem[Aird and Salje(1998)]{Aird1998}
Aird, A.; Salje, E.K.H.
\newblock Sheet superconductivity in twin walls: Experimental evidence of
  WO$_{3-x}$.
\newblock {\em J. Phys. Condens. Matter} {\bf 1998}, {\em
  10},~L377--L380.
\newblock
  doi:{\changeurlcolor{black}\href{https://doi.org/10.1088/0953-8984/10/22/003}{\detokenize{10.1088/0953-8984/10/22/003}}}.

\bibitem[Kopelevich \em{et~al.}(2015)Kopelevich, da~Silva, and
  Camargo]{Kopelevich2015}
Kopelevich, Y.; da~Silva, R.R.; Camargo, B.C.
\newblock Unstable and elusive superconductors.
\newblock {\em Physica C: Supercond. and Its Appl.} {\bf 2015},
  {\em 514},~237--245. 
  doi:{\changeurlcolor{black}\href{https://doi.org/https://doi.org/10.1016/j.physc.2015.02.027}{\detokenize{10.1016/j.physc.2015.02.027}}}.

\bibitem[Reich and Tsabba(1999)]{Reich1999}
Reich, S.; Tsabba, Y.
\newblock Possible nucleation of a 2D superconducting phase on WO single
  crystals surface doped with Na.
\newblock {\em  Eur. Phys. J. B Condens. Matter Complex Syst.} {\bf 1999}, {\em 9},~1--4.
\newblock
  doi:{\changeurlcolor{black}\href{https://doi.org/10.1007/s100510050735}{\detokenize{10.1007/s100510050735}}}.

\bibitem[Shengelaya and M{\"u}ller(2019)]{Shengelaya2019}
Shengelaya, A.; M{\"u}ller, K.A.
\newblock Superconductivity in Oxides Generated by Percolating Electron or Hole
  Bipolarons.
\newblock {\em J. Supercond. Nov. Magn.} {\bf 2019},
  {\em 32},~3--6.
\newblock
  doi:{\changeurlcolor{black}\href{https://doi.org/10.1007/s10948-018-4882-6}{\detokenize{10.1007/s10948-018-4882-6}}}.

\bibitem[Hamdi \em{et~al.}(2016)Hamdi, Salje, Ghosez, and Bousquet]{Hamdi2016}
Hamdi, H.; Salje, E.K.H.; Ghosez, P.; Bousquet, E.
\newblock First-principles reinvestigation of bulk ${\mathrm{WO}}_{3}$.
\newblock {\em Phys. Rev. B} {\bf 2016}, {\em 94},~245124.
\newblock
  doi:{\changeurlcolor{black}\href{https://doi.org/10.1103/PhysRevB.94.245124}{\detokenize{10.1103/PhysRevB.94.245124}}}.

\bibitem[Wijs \em{et~al.}(1999)Wijs, Boer, Groot, and Kresse]{Wijs1999}
Wijs, G.A.d.; Boer, P.K.d.; Groot, R.A.d.; Kresse, G.
\newblock Anomalous behavior of the semiconducting gap in ${\mathrm{WO}}_{3}$
  from first-principles calculations.
\newblock {\em Phys. Rev. B} {\bf 1999}, {\em 59},~2684--2693.
\newblock
  doi:{\changeurlcolor{black}\href{https://doi.org/10.1103/PhysRevB.59.2684}{\detokenize{10.1103/PhysRevB.59.2684}}}.

\bibitem[Wang \em{et~al.}(2011)Wang, Di~Valentin, and
  Pacchioni]{Wang2011_wo3_prb}
Wang, F.; Di~Valentin, C.; Pacchioni, G.
\newblock Semiconductor-to-metal transition in WO${}_{3\ensuremath{-}x}$:
  Nature of the oxygen vacancy.
\newblock {\em Phys. Rev. B} {\bf 2011}, {\em 84},~073103.
\newblock
  doi:{\changeurlcolor{black}\href{https://doi.org/10.1103/PhysRevB.84.073103}{\detokenize{10.1103/PhysRevB.84.073103}}}.

\bibitem[Mehmood \em{et~al.}(2016)Mehmood, Pachter, Murphy, Johnson, and
  Ramana]{Mehmood2016}
Mehmood, F.; Pachter, R.; Murphy, N.R.; Johnson, W.E.; Ramana, C.V.
\newblock Effect of oxygen vacancies on the electronic and optical properties
  of tungsten oxide from first principles calculations.
\newblock {\em J. Appl. Phys.} {\bf 2016}, {\em 120},~233105. 
  doi:{\changeurlcolor{black}\href{https://doi.org/10.1063/1.4972038}{\detokenize{10.1063/1.4972038}}}.

\bibitem[Migas \em{et~al.}(2010)Migas, Shaposhnikov, Rodin, and
  Borisenko]{Migas2010_1}
Migas, D.B.; Shaposhnikov, V.L.; Rodin, V.N.; Borisenko, V.E.
\newblock Tungsten oxides. I. Effects of oxygen vacancies and doping on
  electronic and optical properties of different phases of WO{$_3$}.
\newblock {\em J. Appl. Phys.} {\bf 2010}, {\em 108},~093713.
\newblock
  doi:{\changeurlcolor{black}\href{https://doi.org/10.1063/1.3505688}{\detokenize{10.1063/1.3505688}}}.

\bibitem[Walkingshaw \em{et~al.}(2004)Walkingshaw, Spaldin, and
  Artacho]{Walkingshaw2004}
Walkingshaw, A.D.; Spaldin, N.A.; Artacho, E.
\newblock Density-functional study of charge doping in ${\mathrm{WO}}_{3}$.
\newblock {\em Phys. Rev. B} {\bf 2004}, {\em 70},~165110.
\newblock
  doi:{\changeurlcolor{black}\href{https://doi.org/10.1103/PhysRevB.70.165110}{\detokenize{10.1103/PhysRevB.70.165110}}}.

\bibitem[Tosoni \em{et~al.}(2014)Tosoni, Di~Valentin, and
  Pacchioni]{Tosoni2014}
Tosoni, S.; Di~Valentin, C.; Pacchioni, G.
\newblock Effect of Alkali Metals Interstitial Doping on Structural and
  Electronic Properties of WO{$_3$}.
\newblock {\em  J. Phys. Chem. C} {\bf 2014}, {\em
  118},~3000--3006.
\newblock
  doi:{\changeurlcolor{black}\href{https://doi.org/10.1021/jp4123387}{\detokenize{10.1021/jp4123387}}}.

\bibitem[Huda \em{et~al.}(2008)Huda, Yan, Moon, Wei, and Al-Jassim]{Huda2008}
Huda, M.N.; Yan, Y.; Moon, C.Y.; Wei, S.H.; Al-Jassim, M.M.
\newblock Density-functional theory study of the effects of atomic impurity on
  the band edges of monoclinic ${\text{WO}}_{3}$.
\newblock {\em Phys. Rev. B} {\bf 2008}, {\em 77},~195102.
\newblock
  doi:{\changeurlcolor{black}\href{https://doi.org/10.1103/PhysRevB.77.195102}{\detokenize{10.1103/PhysRevB.77.195102}}}.

\bibitem[Migas \em{et~al.}(2010)Migas, Shaposhnikov, and
  Borisenko]{Migas2010_2}
Migas, D.B.; Shaposhnikov, V.L.; Borisenko, V.E.
\newblock Tungsten oxides. II. The metallic nature of Magnéli phases.
\newblock {\em J. Appl. Phys.} {\bf 2010}, {\em 108},~093714.
\newblock
  doi:{\changeurlcolor{black}\href{https://doi.org/10.1063/1.3505689}{\detokenize{10.1063/1.3505689}}}.

\bibitem[Magn\'{e}li(1949)]{Magneli1949}
Magn\'{e}li, A.
\newblock Crystal structure studies on beta-tungsten oxide.
\newblock {\em Arkiv Kemi} {\bf 1949}, {\em 1},~513--523.

\bibitem[elk()]{elk}
{The Elk Code}.  Available online:
\newblock \url{http://elk.sourceforge.net/} (accessed on 01 December 2020). 

\bibitem[{Perdew} \em{et~al.}(1996){Perdew}, {Burke}, and
  {Ernzerhof}]{jperdew96}
{Perdew}, J.P.; {Burke}, K.; {Ernzerhof}, M.
\newblock {Generalized Gradient Approximation Made Simple}.
\newblock {\em Phys. Rev. Lett.} {\bf 1996}, {\em 77},~3865--3868.
\newblock
  doi:{\changeurlcolor{black}\href{https://doi.org/10.1103/PhysRevLett.77.3865}{\detokenize{10.1103/PhysRevLett.77.3865}}}.

\bibitem[Hinuma \em{et~al.}(2017)Hinuma, Pizzi, Kumagai, Oba, and
  Tanaka]{Hinuma2017}
Hinuma, Y.; Pizzi, G.; Kumagai, Y.; Oba, F.; Tanaka, I.
\newblock Band structure diagram paths based on crystallography.
\newblock {\em Comput. Mater. Sci.} {\bf 2017}, {\em 128},~140--184.
\newblock
  doi:{\changeurlcolor{black}\href{https://doi.org/https://doi.org/10.1016/j.commatsci.2016.10.015}{\detokenize{10.1016/j.commatsci.2016.10.015}}}.

\bibitem[Korshunov \em{et~al.}(2020)Korshunov, Nekrasov, Pavlov, and
  Slobodchikov]{Korshunov2021}
Korshunov, M.; Nekrasov, I.; Pavlov, N.; Slobodchikov, A.
\newblock Band structure for tungsten oxide W$_{20}$O$_{58}$ with ideal
  octahedra.
\newblock {\em Pis'ma v ZhETF} {\bf 2020}, {\em 113},~in print.

\bibitem[Hirschfeld \em{et~al.}(2011)Hirschfeld, Korshunov, and
  Mazin]{HirschfeldKorshunov2011}
Hirschfeld, P.J.; Korshunov, M.M.; Mazin, I.I.
\newblock Gap symmetry and structure of Fe-based superconductors.
\newblock {\em Rep. Prog. Phys.} {\bf 2011}, {\em 74},~124508.

\bibitem[Kontani and Onari(2010)]{Kontani}
Kontani, H.; Onari, S.
\newblock Orbital-Fluctuation-Mediated Superconductivity in Iron Pnictides:
  Analysis of the Five-Orbital Hubbard-Holstein Model.
\newblock {\em Phys. Rev. Lett.} {\bf 2010}, {\em 104},~157001.
\newblock
  doi:{\changeurlcolor{black}\href{https://doi.org/10.1103/PhysRevLett.104.157001}{\detokenize{10.1103/PhysRevLett.104.157001}}}.

\bibitem[Korshunov(2014)]{Korshunov2014eng}
Korshunov, M.M.
\newblock Superconducting state in iron-based materials and spin-fluctuation
  pairing theory.
\newblock {\em Physics-Uspekhi} {\bf 2014}, {\em 57},~813--819.
\newblock
  doi:{\changeurlcolor{black}\href{https://doi.org/10.3367/UFNe.0184.201408h.0882}{\detokenize{10.3367/UFNe.0184.201408h.0882}}}.

\end{thebibliography}
\end{document}